\documentstyle[12pt]{article}

\def\be{\begin{equation}}
\def\en{\end{equation}}

\begin{document}
\begin{titlepage}
\baselineskip = 25pt
\begin{center}
{\Large\bf SECONDARY GRAVITATIONAL ANISOTROPIES IN OPEN UNIVERSES}

\vspace{.5 cm}
{\bf Vicent Quilis
and Diego S\'aez}\\
\small
Departamento de Astronom\'{\i}a y 
Astrof\'{\i}sica. Universidad de Valencia.\\
46100 Burjassot (Valencia), Spain.\\
\footnotesize
e-mail: diego.saez@uv.es, vicente.quilis@uv.es\\
\end{center}

\vspace {2. cm}
\normalsize
\begin{abstract}

The applicability of the potential approximation in the case
of open universes is tested. Great Attractor-like structures are
considered in the test.
Previous estimates of the Cosmic Microwave background anisotropies
produced by these structures are 
analyzed and interpreted. The anisotropies corresponding to
inhomogeneous 
ellipsoidal models are also computed. It is proved 
that, whatever the spatial symmetry may be, 
Great Attractor-like objects with extended cores  
(radius $\sim 10h^{-1}$),
located at redshift $z=5.9$ in an open universe with density
parameter $\Omega_{0}=0.2$,  produce secondary gravitational 
anisotropies of the order of $10^{-5}$ on angular scales of
a few degrees. 
The amplitudes and angular scales of the estimated anisotropy  
decrease as the Great Attractor size decreases.
For comparable normalizations and compensations, the 
anisotropy produced by spherical realizations is found to be 
smaller than that of ellipsoidal models.
This anisotropy appears to be an integrated effect
along the photon geodesics.
Its angular scale is much greater than that subtended 
by the Great Attractor itself. This is understood easily
taking into account that the integrated effect is produced by the
variations of the gravitational potential, which seem to be 
important in large regions subtending angular scales of various degrees.
As a result of the large size of these regions, 
the spatial curvature of the universe becomes important and, 
consequently, 
significant errors ($\sim 30$ per cent) arise in estimations based on the 
potential approximation.
As it is emphasized in this paper, 
two facts should be taken into account carefully in some
numerical estimates of
secondary gravitational anisotropies in open universes: (1)
the importance of scales much greater than those
subtended by the cosmological structures themselves, and (2)
the compatibility of the potential approximation with the
largest scales. 

\end{abstract}

{\em Subject headings:} cosmic microwave background---cosmology:
theory---large-scale structure of the universe

\end{titlepage}

\section{INTRODUCTION}

Recently, the Tolman-Bondi solution of the Einstein equations 
(Tolman 1934; Bondi 1947) was
used in order 
to estimate the Cosmic Microwave Background (CMB) anisotropies
produced by Great Attractor-like (GAL) structures 
(Panek 1992; S\'aez, Arnau \& Fullana 1993;
Arnau, Fullana \& S\'aez 1994; Fullana, S\'aez \& Arnau 1994;
S\'aez, Arnau \& Fullana 1995). These structures were
placed at a wide range
of redshifts. The most interesting results appeared in the case
of an open universe with density parameter 
$\Omega_{0} \leq 0.4$. In this case, it was proved that GAL
objects placed at redshifts
between 2 and 30 produce CMB anisotropies having  
an amplitude of the order of $10^{-5}$ and
an angular scale of a few degrees. For $\Omega_{0}=0.2$, 
the maximum amplitude corresponds to a redshift 
$z \sim 5.9$. What is the origin of these secondary anisotropies?. 

Calculations 
based on the Tolman-Bondi solution seem not to be 
appropriate in order to answer the above question. 
These calculations are based on an numerical integration along
the null geodesics of the Tolman-Bondi spacetime. Such a
general and abstract method lead to accurate results, but it difficults the
splitting of different posible effects contributing to the 
total 
anisotropy and, consequently, the origin of the predicted effect 
does not become clear.

In the case $\Omega_{0}=0.2$, it was verified that the density
contrast of normalized GAL structures is close to unity  at $z=5.9$; namely,
these structures were evolving in the 
mildly nonlinear regime when they 
influenced the CMB photons. This fact suggests the absence of
strong nonlinear effects (see Rees \& Sciama 1968 where some sources
of nonlinear effects are described qualitatively). 
Furthermore, it was also verified that the anisotropy
produced by a GAL object at $z=5.9$ decreases strongly as 
the density parameter increases (Arnau, Fullana \& S\'aez 1994). 
After these considerations,
it seems that we are concerned with anisotropies 
produced by the time variation of the gravitational
potential (which depends on $\Omega_{0}$ strongly). Nevertheless, such an 
interpretation is only a qualitative one. 
How can we obtain a quantitative verification of this interpretation?.
What is the best formalism to do it?.

Let us try to answer these questions after some necessary words
about notation.
Hereafter, $a$ is the scale factor, t is the cosmological time, 
and $\Omega$ is the density parameter. Whatever F may be, 
$F_{0}$, $F_{_{I}}$, and $F_{_{B}}$ stand for the present,
initial and background values of $F$, respectively.
The universe is considered to be open  and its present density 
parameter is fixed to be $\Omega_{0}=0.2$. The reduced Hubble constant
is $h=H_{0}/100$, where $H_{0}$ is the Hubble constant in units of
$Km \ s^{-1}  Mpc^{-1}$. Latin indices run from 1 to 3. Coordinates
$x^{i}$ are pseudo-cartesian. In formulae, 
units are chosen in such a way 
that $8 \pi G = c = 1$, where $G$ and $c$ are the gravitational
constant and the speed of the light, respectively.
Vector $\vec {v}$ stands for 
the peculiar velocity at an arbitrary point P, and $\vec {v}_{r}$ is 
the component of the peculiar velocity in the direction of the line
joining the point P and the  
centre of the GAL object. The energy density and the energy density contrast
are denoted $\rho$ and $\delta$, respectively.

In the case of objects located far from the observer and his last 
scattering surface, the Sachs-Wolfe (1967) effect and the 
Doppler anisotropies produced by peculiar motions are
negligible. On account of this fact,  the anisotropy produced by
these objects 
appears to be an integrated 
effect due to the variations of the gravitational potential
(Mart\'{\i}nez-Gonz\'alez, Sanz \& Silk
1990; Sanz et al. 1996). This anisotropy is given by the 
formula: 
\be
\frac{\Delta T}{T} \sim 
-2\int_{e}^{o} { {\bf \nabla} \phi  (x^{i},t) dx^{i}} 
\sim 2\int_{e}^{o} \frac {\partial{\phi}(x^{i},t)}{\partial{t}} dt  \ ,
\en
\noindent
where $e$ and $o$ stand for emitter and observer,
respectively, $\bf \nabla $ is the gradient operator and
$\phi$ is the potential involved in the line element
\be
ds^2=-(1+2\phi) dt^2 + (1-2\phi)a^2 (1 + \frac {Kr^{2}} {4})^{-2} 
\delta_{ij}dx^idx^j  \ .
\en
\noindent
where $K$ takes on the value $-1$ ($1$, $0$) in the open 
(closed, flat) case 
and $r^{2}=(x^{1})^{2}+(x^{2})^{2}+(x^{3})^{2}$.
This potential satisfies the equation 
\begin{equation}
\nabla^2\phi = \frac{3}{2} H^2 a^2 \Omega \delta \ ,
\end{equation}
\noindent
and, consequently, it can be interpreted as  
the Newtonian gravitational potential.

The integrals involved in Eq.\ (1) are to be carried out along
a null geodesic from the emitter (e) to the observer (o). 
The equations of these geodesics can be derived 
in the background (at zero order). Although this approach was 
initially proposed in the framework of the flat case ($K=0$, $\Omega_{0}=1$),
Sanz et al. (1996) suppose 
it also valid for open universes
in the case of systems having sizes much smaller
than the curvature scale. This validity will be tested at the same time
that we try to answer the following questions:
can we apply the potential approximation in the case of GAL
objects evolving in open universes?. Do we expect some 
errors in the results?. A few considerations about sizes are
necessary in order to consider these questions.

Calculations based on the exact Tolman-Bondi 
solution prove that,  in the case $z=5.9$, $\Omega_{0}=0.2$, 
the secondary anisotropy produced by a GAL  
object has an unexpected angular scale of a few degrees.
This large angular scale is greater than the
angular scale subtended by the GAL itself. This fact is not
surprising, at least, if the resulting anisotropy is 
produced by the gradients of the gravitational
potential. In such a case, 
the region affected by a
significant gravitational potential
is larger than that having a relevant density contrast
(see Fullana, S\'aez \& Arnau 1994). 
At $z=5.9$, scales subtending an angle
between $6$ and $10$ degrees are spatial scales between $210 \ Mpc$
and $350 \ Mpc$, while the curvature scale at the same redshift is
$\sim 970 \ Mpc$. This means that the regions in which the
CMB photons are influenced by the GAL structure is 
between $20$ and $30$ per cent of the horizon scale and,
consequently, the curvature could be relevant and the use of 
Eqs (1)-(3) could leads to significant errors. 
This suspicion will be confirmed by explicit calculations
(see below). This fact is crucial in order to do further
applications --including GAL objects-- of general methods  
for the estimation of secondary gravitational anisotropies 
in open universes (see Section 4).
Among these methods, the numerical approach used by
Tului, Laguna \& Aninos (1996) and the estimates based on
spectra due to Sanz et al. (1996) deserve special attention.  

Calculations based on an exact solution of the Einstein equations
--as the Tolman-Bondi one-- do not involve any gauge or approximating
condition; hence, results are theoretically confident and, consequently,  
if the results obtained from the Tolman-Bondi solution
and from the equations (1)-(3) become comparable, we would have
given strong support to the following ideas: (1) the 
secondary significant effect calculated with
Tolman-Bondi solution is a consequence of the
time variations of the gravitational potential, (2) 
the approach based on Eqs. (1)-(3) applies in the open case 
up to a certain level of accuracy, and (3) calculations based on the 
Tolman-Bondi solution are both theoretically confident and
well performed.  

The main limitation of the estimations based on the 
Tolman-Bondi solution
is expected to be a consequence of spherical symmetry. 
The time evolution of a given
structure and, consequently, the time variations of its gravitational 
potential are conditioned by the symmetry. Hence, according to Eq (1),
the anisotropy could be also affected by 
spherical symmetry. 

The normalization of the Great Attractor is also
strongly affected by this symmetry; in fact, what we know about this
structure is the peculiar velocity that 
it produces on the sourrounding galaxies;
in particular, following (Lynden-Bell et al. 1988), the 
Great Attractor produces a 
peculiar velocity $\mid \vec {v}_{r0} \mid =570 \pm 60 Km \ s^{-1}$ 
at the radial distance 
$R_{0} \sim 43h^{-1} \ Mpc$ corresponding to the distance from Local Group
to the Great Attractor centre. The peculiar velocity field depends on the
distance to the centre, $R_{0}$, as it 
corresponds to the velocity field created by an overdensity
with a density contrast proportional to $1/R_{0}^{2}$. These data
and the size of the core radius 
were used in previous papers for normalizing GAL objects.
It is evident that GAL structures as an 
elongated ellipsoid or a pancake
would produce different peculiar velocities at different points
located at $43 h^{-1} \ Mpc$ from the symmetry centre.
The orientation of the structure 
with respect to the line of sight is crucial to calculate 
$\mid \vec {v}_{r0} \mid$ at the observer position.
In the absence of spherical symmetry, a certain velocity, e.g.   
$\mid \vec {v}_{r0} \mid = 500 \ Km/s$,
can be produced by objects having different masses 
and sizes (depending on the orientation), which would produce 
different effects on the CMB. 
Ellipsoidal homogeneous models were used previously by
Atrio-Barandela \& Kashlinsky
(1992) and Chodorowski (1994) to study the anisotropies produced by 
pancake-like structures. The ellipsoids considered in this paper are
inhomogeneous.  

This paper is organized as follows. In Section 2, spherical,
elongated, and flattened GAL structures are described. All of them are 
normalized according to Lynden-Bell et al. (1988).
Anisotropies are calculated in Section 3. Calculations are based on
the potential approximation; namely, on Eqs (1)--(3). First, results
corresponding to the spherically symmetric case are compared to
those obtained with the Tolman-Bondi solution for the same model.
In this way, the approach based on Eqs. (1)--(3) is tested
and the interpretation of the resulting anisotropy --as produced
by a varying gravitational potential-- is verified. The effects of
the deformation with respect to spherical symmetry are then 
analyzed. Main conclusions and a general discussion are
summarized in Section 4.

\section{GREAT ATTRACTOR-LIKE MODELS}

Whatever the spatial symmetry of the GAL structure may be, 
the following normalization condition is assumed:
{\em At present time, 
there are points located at $43h^{-1} \ Mpc$  from the 
GAL centre where  $\mid \vec {v}_{r0} \mid = 500 \ Km/s$}. 
We are interested in this low 
velocity a little smaller than the minimum velocity claimed by
Linden-Bell et al (1988) because such a velocity allows us to obtain 
lower limits to the anisotropy produced by admissible GAL structures. 

In what is called our spherical main (SM) model, initial conditions are
chosen in the same way as in previous calculations based on 
the Tolman-Bondi solution.
These conditions are set at redshift $z_{i}=1000$. As in 
Arnau, Fullana \& S\'aez (1994), the initial profile 
of the density contrast is 
\begin{equation}
\delta_{_{I}} = \frac 
{\rho_{_{I}} - \rho_{_{BI}}} {\rho_{_{BI}}} =
\frac{\epsilon_{1}}{1 +
\left( R / R_{1} \right)^{2}} + \frac{\epsilon_{2}}{1 +
\left( R / R_{2} \right)^{2}}  \ ,
\end{equation}
where $R$ is a radial coordinate and the parameters
$\epsilon_{1}$ ($\epsilon_{2}$) and $R_{1}$ ($R_{2}$) 
set the amplitude and the
size of a central overdensity (surrounding underdensity). The
conditions $\epsilon_{1} > 0$, $\epsilon_{2} < 0$,
$\epsilon_{1} > \epsilon_{2}$, and $R_{2} > R_{1}$
must be satisfied. 
The initial peculiar velocity is that corresponding to the above density 
profile in the case of vanishing nongrowing modes. In the SM model, 
the above parameters take on the values  
$\epsilon_{1}=5.42 \times 10^{-3}$, $\epsilon_{2}=-6.7 \times 10^{-4}$,
$R_{1}=4.21 \times 10^{-2} h^{-1}  Mpc$, and
$R_{2}=1.2 \times 10^{-1} h^{-1}  Mpc$. In this case, the 
above normalization condition  
is satisfied and 
compensation is achieved at the last scattering surface 
of an observer placed at the GAL centre. The present density
profile of the resulting structure was given in 
Arnau, Fullana, S\'aez (1994,
see curve C2 of Fig.\ 2). From this Figure, it follows that
the contrast reduces to one-half of the central maximum value 
at $9h^{-1} \ Mpc$. This profile shows a strong resemblance with 
the Great Attractor described by Linden-Bell et al. (1988).
In spite of the fact that the exact compensation takes place very far from 
the symmetry centre, the present density contrast decreases 
fastly and, consequently, the energy density is negligible at
distances much smaller than that of exact compensation.  
The SM model plays the role of an important reference,
which has been well studied without approximating conditions; namely,
with an exact cosmological solution of Einstein equations

In order to model GAL structures without spherical symmetry,
the following initial density contrast has been
chosen:
\begin{equation}
\delta_{_{I}} = \frac {\epsilon_{1}}{1 +
\sum_{i=1}^{3} X_{i}^{2} / A_{1i}^{2}} 
+ \frac {\epsilon_{2}}{1 +
\sum_{i=1}^{3} X_{i}^{2} / A_{2i}^{2}} \ ,
\end{equation}
where $X_{i}=ax^{i}$. The quantities 
$A_{1i}$ and $A_{2i}$ satisfy the following relations:
\begin{equation}
A_{2i}=mA_{1i} \ ,  m>1 \ ,
\end{equation}
\begin{equation}
A_{11}=A_{12}=A_{1} \ , A_{13}=nA_{1} \ , 
\end{equation}
where $A_{1}$, $m$ and $n$ are free parameters defining the 
shape of the 
isodensity surfaces, which are ellipsoids.
For $n=1$, the profile (5) is spherically
symmetric (S profiles).
In the case $n > 1$, the isodensity surfaces 
are ellipsoids 
elongated (E profiles) along the $x_{3}$ direction.   
For $n < 1$, these surfaces are 
pancake-like (P profiles) 
ellipsoids. Directions $x_{1}$ and $x_{2}$ are undistinguisable.
The structure is either elongated or flattened along the 
$x_{3}$ direction. These criteria about the names and characteristics
of the axis must be taken into account carefully in order to 
imagine the orientations of the ellipsoids defined below.

The profile (5) has two appropriate features: (1) In the
spherically symmetric limit $n=1$, if one takes  
$A_{1i}=R_{1}$, $A_{2i}=R_{2}$ for 
$i=1,2,3$, it 
reduces to the profile (4), which seems to have
an admissible dependence on $R$, 
in agreement with Linden-bell et al. suggestions based
on the peculiar velocity field (1988), 
and (2) in spite of the multidimensional 
features of the mass distribution, the gravitational potential corresponding
to this profile can be calculated after performing a one-dimensional   
integral. From Binney and Tremaine
(1987) and references cited therein, it can be proved easily
that, inside the density distribution, 
the gravitational potential is of the form
\begin{equation}
\phi (x^{i}) = - \frac {1} {8} \frac {B_{2} B_{3}} {B_{1}}
\int_{0}^{\infty} \frac { \psi (\infty) - \psi (m) }
{[(B_{1}^{2} + \tau)+(B_{2}^{2} + \tau)+(B_{3}^{2} + \tau)]^{1/2}}
d \tau
\end{equation}
\noindent
where $\psi (m)$ is defined as follows
\begin{equation}
\psi (m) = \int_{0}^{m^{2}} \delta_{_{I}}(m^{2}) d(m^{2})
\end{equation}
and 
\begin{equation}
m^{2} = B_{1}^{2} \sum_{i=1}^{3} \frac {(x^{i})^{2}} {B_{i}^{2} + \tau}
\end{equation}
The parameter $\tau$ defines isopotential surfaces and
the quantities $B_{i}$ are proportional to $A_{i}$ and define the
boundary of the density distribution. We have taken a large enough 
value of the proportionality constant between $B_{i}$ and $A_{i}$;
thus, we are always concerned with the potential inside the distribution
and Eq. (8) applies.
In the asymmetric case, the gravitational potential  
is calculated   
by using the following formula: 
\begin{equation}
\phi(x^{i},t)=\frac {3}{2} H^{2} a^{2} \Omega D \phi(x^{i}) \ ,
\end{equation}
where $\phi(x^{i})$ is given by Eq.\ (8), and
$D$ is the growing mode of the density contrast given by the following 
equations (Peebles, 1980): 
\begin{equation}
D = 1 + \frac {3} {y} + \frac {3(1+y)^{1/2}} {y^{3/2}}
ln [(1+y)^{1/2} - y^{1/2}] \ ,
\end{equation}
\begin{equation}
y = a \frac {H_{0}(1- \Omega_{0})^{3/2}} {\Omega_{0}} \ .
\end{equation}
The normalization of the GAL objects is achieved by using the formula
\begin{equation}
v^{j}=- \frac {2} {3H \Omega D} \frac {dD} {da} 
\frac {\partial \phi(x^{i},t)} {\partial x^{j}} 
\end{equation}
The potential given by Eq. (11) can be obtained by using 
Eq. (3) and the 
linearized density contrast $\delta=D \delta_{_{I}}$.
The application of the above linearized equations
requires discussion. Since normalization is based 
on the estimate of a present velocity and the Great Attractor 
is not a linear
structure at present (see Fullana, S\'aez \& Arnau 1994, for an
estimate of the amplitude of the density contrast at $t_{0}$),
the following question arises:
can we use Eqs. (11)--(14) for normalization?.
In order to answer this question we have developed a test.
In the spherically symmetric case, the velocity predicted by
Eqs. (11)--(14) has been compared with that calculated with
the Tolman Bondi solution (exact nonlinear estimate). 
The resulting present velocity
is $523 \ Km/s$ to be compared with the velocity of $500 \ Km/s$
given by Tolman-Bondi calculations; hence, the relative error given
by the above approach --in the present peculiar velocity and,
consequently, in
the present spatial gradients of the gravitational 
potential-- is $\sim 5 \%$.
This is in 
agreement with the known ansatz that velocities keep linear
after the density contrast becomes nonlinear. 
Another important question is: can we use Eqs. (11)--(14) plus
Eq. (1) in our computations of secondary gravitational anisotropies
produced by GAL objects ($z=5.9$, $\Omega_{0}=0.2$)?.
According to Eq. (1), these anisotropies depend
on the gradients
of the gravitational potential along the photon null 
geodesics. The most relevant gradients are those calculated 
near the GAL object --at redsifht $\sim 5.9$-- when the 
amplitude of the density contrast
was close to unity. These gradients are calculated very near 
the linear regime and,
consequently, their errors should be smaller than those of the present
gradients estimated above (a few per cent).
This means that only small errors should appear in our 
estimations based on the Potential Approximation plus 
Eq. (11). This is also confirmed by some nonlinear estimates
presented below. 

In order to normalize E and P profiles according to
the criterium described above, the orientation 
of the GAL structure is crucial. 
This is because the peculiar velocity at the observer position
depends on this orientation strongly.
For each profile, two limit orientations   
are considered in which $\mid \vec {v}_{r0} \mid$ takes on its
limit values. In the first (second) case, the line joining
the observer and the GAL centre is parallel (orthogonal)
to the
$x_{3}$ axis. Thus, four normalizations are distinguised. Hereafter,
these normalizations are denoted EP, EO, PP and PO. The first letter
indicates the type of profile (Elongated or Pancake-like ellipsoids),
while the second letter tell us in an evident manner whetter the
line of sight of the GAL centre is either parallel (P) or orthogonal (O)
to the $x_{3}$ axis. 

Table 1 shows the values of the parameters defining the 
initial density contrasts 
of the Great Attractor realizations studied in detail  
in this paper. In all the cases, the values of the parameter $m$
is $2.85$. This value is that of the SM model used as a reference. 
For each pair ($n$, $A_{1}$), the parameters $\epsilon_{1}$ and
$\epsilon_{2}$ have been obtained from the conditions of normalization
and compensation. Compensation is achieved in such a way that it
occurs at a present isodensity surface 
with a semiaxis of $9260h^{-1} \ Mpc$. This is the
distance from an observer located at the Great Attractor centre to
his last scattering surface at $z=1000$. In the spherically
symmetric case, this compensation reduces to that used 
in the Tolman-Bondi treatment of the SM model.
 
\section{RESULTS}

All the selected GAL structures
are located far from the last scattering surface and, 
consequently, they produce negligible
temperature fluctuations and Doppler shifts on this 
surface; in which,
the temperature is assumed to be constant. Furthermore,
all the GAL objects are also placed far from the observer and,
consequently, they produce negligible 
peculiar velocities at the observer
position. This means that the Doppler kinematic dipole 
and quadrupole produced by the peculiar velocity of the observer
are negligible.
Finally, the Sunyaev-Zeldovich effect (see Sunyaev and
Zel'dovich 1980 for a review) is not considered
at all; hence, we are estimating a pure gravitational effect 
produced far from the last scattering surface and,
consequently, we are concerned with secondary gravitational
anisotropies.  

In the general case, for each normalization, the resulting 
anisotropy depends on both the location of the GAL centre and
the orientation of its axis. This orientation corresponds to a
structure located at redshift $z=5.9$ and, consequently, it
is absolutely independent
on that considered for normalization, which corresponds to
another object located at $43h^{-1} \ Mpc$ from the observer.
For each normalization, two limit orientations -- at $z=5.9$ -- 
are considered.
These orientations correspond to the cases in which the line
of sight of the GAL centre is Parallel (P) and Orthogonal (O) 
to the $x_{3}$ axis. The following notation is used:
For a given normalization, for example EO, two estimations of
the anisotropy are presented, which correspond to the orientations
O and P defined above. Thus, we distinguish eight cases: EPP, EPO,
EOO, EOP,  POO, POP, PPP, PPO.

In the spherically symmetric case, the CMB anisotropy produced by
a certain GAL structure has been computed by using three different
codes. One of them is based on the Tolman-Bondi solution 
(see Arnau Fullana \& S\'aez 1994), the
second one uses a linearized approach based on Eqs. (1)-(3) and (11) 
and, the third code uses Eqs. (1)-(3) and the density contrast
$\delta=D \delta_{_{I}}+D^{2}(\frac {5} {7} \delta_{_{I}}^{2} +
{\bf \nabla} \delta_{_{I}} \cdot {\bf \nabla} \Phi +
\frac {2} {7} \Phi,_{ij} \Phi^{,ij})$ with ${\bf \nabla}^{2} \Phi = 
\delta_{_{I}}$
(see Sanz et al. 1996 for comments and references). This contrast includes
a second order term.
All the codes numerically
compute the temperature T of the microwave background 
as a function of the
observation angle $\psi$; this is the angle formed by the 
line of sight and the line
joining the observer and the inhomogeneity centre.
In order to facilitate comparisons with 
Arnau, Fullana \& S\'aez (1994),
the function $T(\psi)$ is then used to calculate the mean temperature
$<T> \ = \ (1/2) \ \int^{\pi}_{0} T(\psi) \ sin\psi \ d\psi$
and the total temperature
contrast $\delta_{_{GRAV}}(\psi) \ = \ [T(\psi) \ - \ <T>] \ / \ <T>$
and, finally, 
second-order radial
differences at the angular scale $\alpha=8.1^{\circ}$
are calculated, where $\alpha$
stands for the angle between two observational directions. These
second order differences are defined as follows
\begin{equation}
(\Delta T/T)_{8.1}(\psi)= \frac {1} {2} \{ [ \delta_{_{GRAV}}(\psi)
- \delta_{_{GRAV}} (\psi - 8.1)] - [ \delta_{_{GRAV}}(\psi + 8.1)
- \delta_{_{GRAV}} (\psi)] \}
\end{equation}

Figure 1 shows the second order differences $(\Delta T/T)_{8.1}(\psi)$
corresponding to the SM model of reference.
The dashed line displays the results given by calculations based on
the Tolman-Bondi solution, while the solid line has been obtained from
Eqs. (1)-(3) and (11). For these two lines, 
the relative error in the amplitudes appears to be 
$\sim 30$ per cent. The angular scales are similar.
The errors appearing as a result of the use of Eq. (11)
--linear estimation of the gravitational potential-- are expected to be 
smaller than a few per cent (see Section 2).
This expectation about the smallness of the errors produced
by linearization is confirmed by the comparison of the
solid and dotted lines, which have been obtained
from density contrasts approximated up to first and second order, 
respectively. 
These two lines are almost undistinguishable in the 
Figure. An appropriated zoom has been included to display 
the differences. The relative difference between the amplitudes
is $2.3$ \% and, consequently, the effect of the nonlinear
term of the density contrast cannot be the main source of the $\sim 30$ per
cent error mentioned above. The main part of this 
error should be due to the presence of large spatial
scales leading to significant curvature effects.                                      
Are these spatial scales associated to the density contrast
as a result of our compensations at large distances from the symmetry
centre?. In order to test this possibility, compensations 
at much smaller distances from the symmetry center
have been performed and errors near $30 \%$ have been obtained 
in all the cases. This means that the large scales involved in the
problem correspond to the large regions where the
gravitational potential is significant.  

The above discussion about the nonlinear effect justifies the use
of the linear approach in the following applications.

The size of spherical GAL structures has been varied maintaining
the model of Section 2 and its normalization. In order to vary the size, 
the value of $A_{1}$ corresponding to the SM model of Figure 1
(first row of Table 1)  has been multiplied by
the factor $\xi$, while parameters $m$ and $n$ have not been altered. 
The parameter $\xi$, which set
the GAL size, has been varied in the interval [0.5,2]. 
The anisotropy produced by each of the resulting structures has been
estimated.
Figure 2 displays the relation between the amplitude of the 
$(\Delta T/T)_{8.1}(\psi)$ differences and the size of
the GAL structure defined by factor $\xi$. 
The relation $(\Delta T/T)_{8.1}(0)= 4.947 \times 10^{-6} -
2.439 \times 10^{-6} \xi + 1.979 \times 10^{-5} \xi^{2}$ fits
very well the numerical estimates (stars).
It is noticeable
that the CMB anisotropy decreases as the GAL size decreases.
This means that a too concentrated GAL structure 
would not produce any significant effect. Linden-Bell, et al (1988)
claimed that the Great Attractor core has a radius 
$R_{c} \sim 10h^{-1} \ Mpc$ and a $1/R^{2}$ density profile. In this case,
it would produce anisotropies at the level $10^{-5}$; nevertheless,
more observational data have been and are being 
obtained (Kraan-Korteweg, Woudt \& Henning, 1996). 
Future conclusions about the size and mass of the
Great Attractor would be important in order to get a definitive 
estimation of its contribution to 
the CMB anisotropy in open universes.

In the spherically symmetric case, all the planes containing
the line of sight of the symmetry centre are equivalent.
In the absence of spherical symmetry, we 
present $(\Delta T/T)_{8.1}(\psi)$ differences in two planes.
Each of these planes is  
generated by the line of sight of the symmetry centre
and one of the ellipsoid axis perpendicular to this line. 
They are two orthogonal planes.  
For a fixed normalization, 
if the axis perpendicular to the line of sight are the
undistinguishible axis $x_{1}$
and $x_{2}$, the second order differences corresponding to both planes 
should coincide. On the contrary,
for the pairs ($x_{1}$, $x_{3}$) and ($x_{2}$, $x_{3}$),
the resulting differences are expected to be different. The differences 
between the results corresponding to both planes are due
to the asymmetry of the GAL object which has been located 
and orientated at $z=5.9$.

Figure 3 shows the second order differences $(\Delta T/T)_{8.1}(\psi)$
for the eight cases defined in Section 2. Table 2 gives the values of the
maxima, $(\Delta T/T)_{8.1}(0)$, 
appearing in the curves of this Figure. The chosen cases 
correspond
to various profiles, normalizations and orientations of the
symmetry axis at $z=5.9$.
In each panel, the type of profile (first letter inside 
the panel: E or P), the normalization 
(second letter: P or O) and the orientation at $z=5.9$ 
(third letter P or O) have been fixed. 
Continuous and dashed lines display second
order differences in each of the two orthogonal planes 
defined above. As expected, when the axis perpendicular 
to the line of sight are $x_{1}$ and $x_{2}$ (the third letter
is a P), both curves 
coincide. In the remaining cases, the continuous and dashed 
lines are different but rather similar. 
The two top panels (fourth level in the Figure) 
correspond to E profiles normalized in 
the same way, but having a different orientation at $z=5.9$. The
amplitude of the left (right) panel is $\sim 5.7 \times 10^{-5}$
($\sim 4.8 \times 10^{-5}$); hence, the orientation at $z=5.9$ is not very 
important. It leads to deviations of about $15$ per cent. The
same conclusion is obtained from comparisons of the left and
right panels in the third, the second and the
first level of panels (see Table 2 for the values of the amplitudes).
Comparisons of the panels corresponding to distinct levels show
the importance of the profile and the normalization.
For E (P) profiles, the differences between the third (first)
and the fourth (second) levels
of panels appear as a result of normalization.
We can conclude than the normalization 
--more properly the orientation defining the normalization-- 
is very important. It can modify the resulting anisotropy
by a factor $\sim 2$. As it can be seen in Table 2, 
the smallest (greatest) amplitude appears in the case
EOO (POO) and its value is $\sim 2.5 \times 10^{-5}$
($\sim 7.0 \times 10^{-5}$). In any case, the resulting anisotropy is 
very significant.

In Figure 4, we have considered a EPP case with 
the semi-axis reduced by a factor $1/2$ with respect to those
of the EPP realization of Figure 3. As in the spherically symmetric
case, we can see that the CMB anisotropy has decreased.
The amplitude corresponding to Fig. 4 
is $1.34 \times 10^{-5}$.

\section{CONCLUSIONS AND DISCUSSION}

In this paper, the following elements have been fixed: 
a normalization condition
for GAL structures (first
paragraph of Section 2), a form for the initial
density profile, a compensation distance, 
an open background ($\Omega_{0} = 0.2$),
and a redshift for location ($z=5.9$). The resulting GAL
structures are very similar to that described by Linden-Bell et al. (1988).

The spherically symmetric case has been analyzed in detail.
The use of both the potential approximation and the Tolman-Bondi
solution has proved that, for $z=5.9$ and $\Omega_{0}=0.2$,
objects having cores with present sizes near a ten of 
Megaparsecs produce significant anisotropies with amplitudes
of $10^{-5}$ on scales of various degrees. 
This is because 
the gradients of the gravitational potential are significant
on spatial scales of a few hundred of Megaparsecs 
(at z=5.9) subtending angular scales
of various degrees; however, 
the angle subtended by the central region where the density 
contrast is significant is 
smaller than the angular scale of the anisotropy.  

Starting from the same initial conditions (first row of Table 1),
computations based on both the Tolman-Bondi solution 
and Eqs (1)--(3) plus (11) have given
comparable anisotropies with very similar
scales and a little different amplitudes. 
It is noticeable that the relative differences between the
resulting amplitudes -- $\sim 0.3$ -- is very similar
to the ratio between the scales where the CMB photons feel 
the variations of the gravitational potential and the curvature scale.
Since the errors due to the use of Eq. (11) should justify only 
a small part of these relative differences, the most important 
part of them should appear
as a result of the application of the potential approximation to
extended regions where the curvature should be taken into
account rigorously.

Finally, in the spherically symmetric case with the profile
(4), it has been proved
that the anisotropy produced by GAL objects depends strongly
on their size. 
This dependence was not pointed out in previous papers 
based on the Tolman-Bondi solution.
In spite of the fact that the normalization has
not been changed, the amplitude of the resulting anisotropy and
its angular scale 
decreases as the size of the object decreases.  

In the case of the pancake-like and ellipsoidal-like GAL structures
with the profile (5), the angular scales have appeared to be very similar
to those of the spherically symmetric case, but the amplitudes 
are rather different. In some cases, the amplitude of ellipsoidal structures 
become magnified by a factor near $3$ with respect to the 
case of spherical objects. Furthermore, it has been verified that: 
(a)  the amplitudes
depend on the orientation of the axis at $z=5.9$, but this dependence 
is weak, (b) the orientation assumed for normalization is very important
because it conditionates the size and mass of the resulting structure and,
consequently, the anisotropy amplitude, and (c)
the amplitude and the angular
scale of the anisotropy depend on the spatial size 
of the structure as it occurs in the spherically symmetric case.
 
The anisotropy produced by all the GAL objects 
of the Universe essentially depends on: the features of the
Great Attractor (normalization and size), the value of
the density parameter and the abundance of this kind of objects.
New observations giving information about some of these 
elements would be necessary  
in order to obtain 
definitive conclusions from Fig. 3. In fact, if the value of the
density parameter is found to be $\Omega_{0}=0.2$ and various 
GAL objects are observed between $z=2$ and $z=30$
(see Arnau, Fullana, S\'aez 1994), data from C.O.B.E. 
satellite and TENERIFE experiment
would rule out various GAL realizations of Fig. 3. In fact, these 
experiments give amplitudes $\sim 10^{-5}$ for angular scales
of a few degrees and, consequently, realizations EPP, EPO, POO and
POP would be inadmissible. According to the same experiments, 
if the Great Attrator is found to 
be an EPP realization and GAL objects are abundant enough, 
the density parameter
$\Omega_{0}=0.2$ is forbidden. On the contrary, if the universe 
is found to be quasi-flat, no conclusions would be obtained 
from Fig. 3 because any GAL objects would produce negligible anisotropies. 

The fact that the angular scales appear to be of various degrees --in
all the cases considered in this paper-- should be taken into
account in other types of usual calculations. Let us discuss this
point in more detail. Suppose that we take a Fourier box where
a certain realization of structures is generated. If we wish to
estimate --numerically-- the anisotropies appearing in a universe filled 
by these boxes, only the effect of structures much smaller than 
the size of the box can be taken into account. This is a well known fact,
which must be analyzed in the case of the secondary
anisotropies given by Eq. (1); in this case, according to our conclusions, 
the box should be much greater than the regions where
the variations of the potential are significant (not greater than 
the density structures). This means that calculations of the effect 
produced by 
various possible realizations of the Great Attractor 
would require too big boxes with a huge size of thousands of
Megaparsecs. 
In calculations based on Eqs. (1)--(3) plus spectra and statistics,
it should be taken into account that: (1) these equations lead to
substantial error in the presence of big structures as the Great
attractor ($\sim 30$ per cent), 
and ($2$) the usual spectra and statistics and specially
their time evolution could require substantial modifications in order
to account for the presence and evolution of GAL objects. 

\vspace{1 cm}

\noindent
{\it Acknowledgements}. This work has been 
supported by the  Generalitat Valenciana (grant GV-2207/94).
Authors thank Dr. M. Fullana for useful discussions.
V. Quilis thanks to the Conselleria d'Educaci\'o i 
Ci\`encia de la Generalitat Valenciana for a fellowship.

\noindent
{\large{\bf References}}\\
\\
\noindent
Arnau J.V., Fullana M.J., \& S\'{a}ez D., 1994, MNRAS, 268, L17\\ 
Atrio-Barandela F. \& Kashlinsky, A., ApJ, 1992, 390, 322\\
Binney, J., \& Tremaine, S., 1987, {\em Galactic Dynamics}.
Princeton Series in Astrophysics. Princeton University Press\\
Bondi H., 1947, MNRAS, 107, 410\\
Chodorowski, M., 1994, MNRAS, 266, 897\\
Fullana M.J., S\'aez D. \& Arnau J.V., 1994, ApJ Supplement,
94, 1\\
Kraan-Korteweg R.C., Woudt P.A., \& Henning P.A., 1996,
Astro-ph/9611099\\
Linden-Bell D., Faber S.M., Burstein D., Davies R.L., Dressler A.,
Terlevich R.J., Wegner G., 1988, ApJ, 326, 19\\ 
Mart\'{\i}nez-Gonz\'alez E., Sanz J.L. \& 
Silk J., 1990, ApJ, 355, L5\\
Panek M., 1992, ApJ, 388, 225\\
Peebles P.J.E., 1980, 
{\em The Large Scale Structure of the Universe}. 
Princenton University Press\\
Rees M.J. \& Sciama D.W., 1968, Nat, 217, 511\\
Sanz J.L., Mart\'{\i}nez-Gonz\'alez E., Cay\'on L.,
Silk J. \& Sugiyama N., 1996, 
ApJ, 467, 485\\
Sachs R.K., \& Wolfe A.M., 1967, ApJ, 147, 73\\
S\'aez D., Arnau J.V. \& Fullana M.J., 1993, MNRAS,
263, 681\\
S\'aez D., Arnau J.V. \& Fullana M.J., 1995, Astro. Lett. Communications,
32, 75\\
Sunyaev R.A. \& Zel'dovich Ya. B., 1980, A\&A. 18, 537\\
Tolman R.C., 1934, Proc. Natl Acad. Sci., 20, 169\\
Tuluie R., Laguna P. \& Anninos P., 1996, ApJ, 463, 15\\

\newpage

\begin{center}
{\bf Figure Captions}
\end{center}

\noindent
{\bf Fig.\ 1.} Second order differences $( \Delta T/T)_{8.1}$ as
functions of the observation angle $\psi$ in degrees for the spherically
symmetric GAL structure corresponding to the first row of Table 1
(SM model).
Continuous, dotted and dashed lines correspond to estimates based on 
a linear approach, a second order approach
and the Tolman-Bondi solution, respectively.
The zoom magnifies the region 
around $\psi=0$. Its axis show the same
quantities as those of the 
main Figure.
 
\vskip 0.5cm
\noindent

\noindent
{\bf Fig.\ 2.} Amplitude of $( \Delta T/T)_{8.1}$ as a function
of the parameter $\xi$ defining the size of the GAL structure.
Stars show the amplitudes numerically estimated and the continuous line
is a good fit to these values. 
\vskip 0.5cm
\noindent

\noindent
{\bf Fig.\ 3.} Same as Fig.\ 1. Each panel correponds to
one of the cases EPP, EPO, EOO, EOP, POO,
POP, PPP, PPO defined in the text. Continuous  and dashed lines 
show $( \Delta T/T)_{8.1}$ differences in the planes
($x_{1},x_{2}$)  and ($x_{1},x_{3}$), respectively. The axis $x_{1}$ 
has the direction of the line of sight. In - - P (- - O) cases,
both curves are identical (different). 
\vskip 0.5cm
\noindent

\noindent
{\bf Fig.\ 4.} Same as in the top left panel of 
Fig.\ 3, but the semiaxis of the ellipsoids have been reduced by
the factor $1/2$. 
\vskip 0.5cm
\noindent

\newpage

\begin{table}
\begin{center}
{\bf Table 1}\\
Realizations of the Great Attractor\\
 \begin{tabular}{cccccc}\\
 
Profile & Orientation & $\epsilon_{1}$ & $\epsilon_{2}$ & 
$A_{1}$ & n \\

 &  &  &  & ($h^{-1} \ Mpc$)  & \\
 &  &  &  &                   & \\

 S & -- & $5.42 \times 10^{-3}$ &  $-6.7 \times 10^{-4}$ &
      $4.21 \times 10^{-2}$ & $1$\\
      
 S & -- & $1.19 \times 10^{-2}$ &  $-1.49 \times 10^{-3}$ &
      $2.1 \times 10^{-2}$ & $1$\\  
      
 E & P & $7.15 \times 10^{-3}$ &  $-9.04 \times 10^{-4}$ &
      $4.21 \times 10^{-2}$ & $2$\\ 
      
 E & O & $4.06 \times 10^{-3}$ &  $-5.28 \times 10^{-4}$ &
      $4.21 \times 10^{-2}$ & $2$\\  
      
 P & O & $5.17 \times 10^{-3}$ &  $-6.54 \times 10^{-4}$ &
      $8.42 \times 10^{-2}$ & $0.5$\\ 
      
 P & P & $3.06 \times 10^{-3}$ &  $-3.97 \times 10^{-4}$ &
      $8.42 \times 10^{-2}$ & $0.5$\\      

 E & P & $1.07 \times 10^{-2}$ &  $-1.34 \times 10^{-3}$ &
      $2.1 \times 10^{-2}$ & $2$\\   
\multicolumn{6}{c}{}\\
\end{tabular}
\end{center}
\end{table} 

\newpage

\begin{table}
\begin{center}
{\bf Table 2}\\
Anisotropies of ellipsoidal Great Attractor-like objects\\
 \begin{tabular}{cccccc}\\
 
CASE & PLANE .. &
$(\Delta T/T)_{8.1}(0)$ & CASE & PLANE & $(\Delta T/T)_{8.1}(0)$\\  

& & & &  &\\

EPP & ($x_{1},x_{3}$) & $5.68 \times 10^{-5}$ & EPP & 
($x_{2},x_{3}$) & $5.68 \times 10^{-5}$\\ 

EPO & ($x_{1},x_{2}$) & $4.86 \times 10^{-5}$ & EPO & 
($x_{1},x_{3}$) & $4.80 \times 10^{-5}$\\

EOO & ($x_{1},x_{2}$) & $2.46 \times 10^{-5}$ & EOO & 
($x_{1},x_{3}$) & $2.52 \times 10^{-5}$\\

EOP & ($x_{1},x_{3}$) & $2.83 \times 10^{-5}$ & EOP & 
($x_{2},x_{3}$) & $2.83 \times 10^{-5}$\\

POO & ($x_{1},x_{2}$) & $6.93 \times 10^{-5}$ & POO & 
($x_{1},x_{3}$) & $7.04 \times 10^{-5}$\\

POP & ($x_{1},x_{3}$) & $6.08 \times 10^{-5}$ & POP & 
($x_{2},x_{3}$) & $6.08 \times 10^{-5}$\\

PPP & ($x_{1},x_{3}$) & $3.37 \times 10^{-5}$ & PPP & 
($x_{2},x_{3}$) & $3.37 \times 10^{-5}$\\

PPO & ($x_{1},x_{2}$) & $3.81 \times 10^{-5}$ & PPO & 
($x_{1},x_{3}$) & $3.76 \times 10^{-5}$\\
\multicolumn{6}{c}{}\\
\end{tabular}
\end{center}
\end{table} 
\newpage

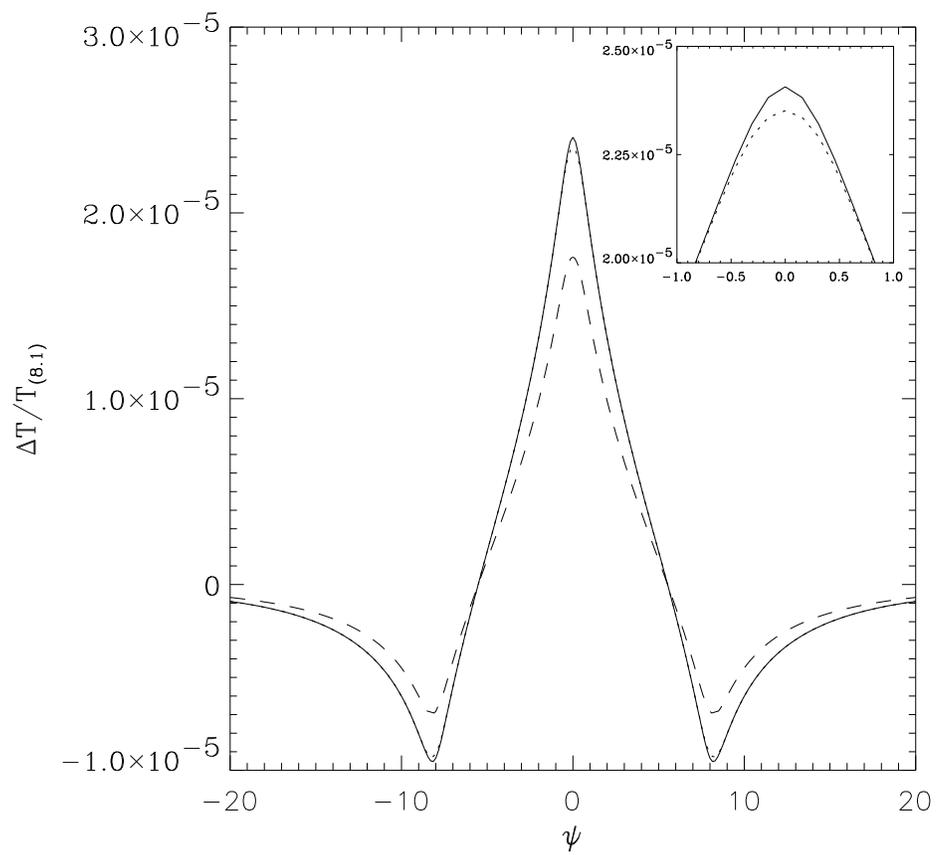
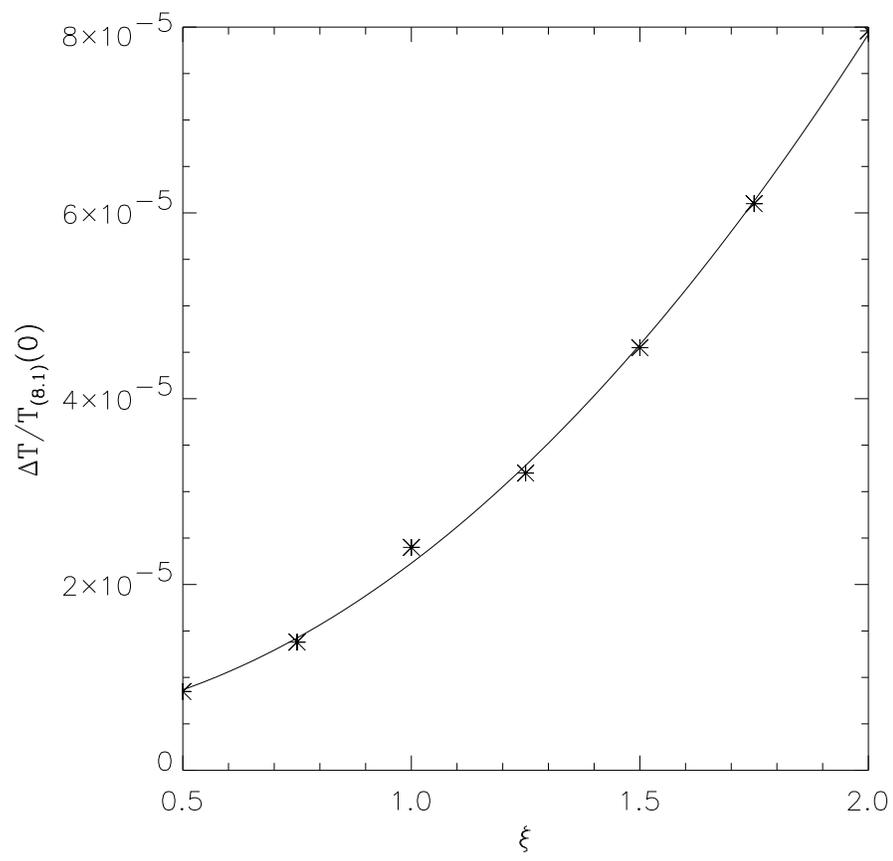
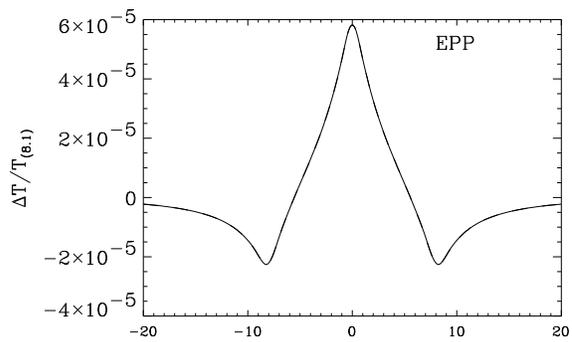
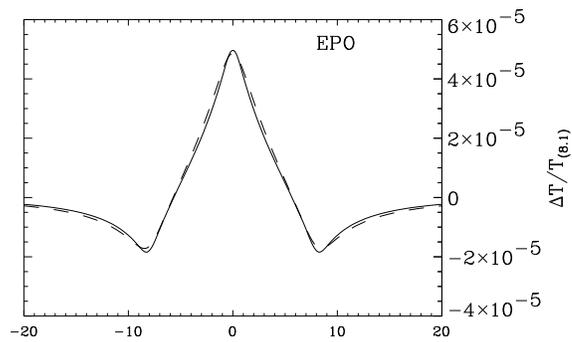
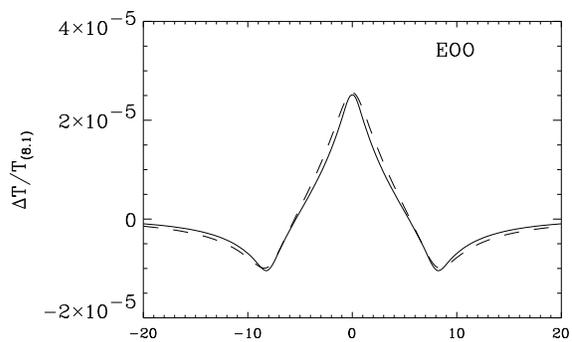
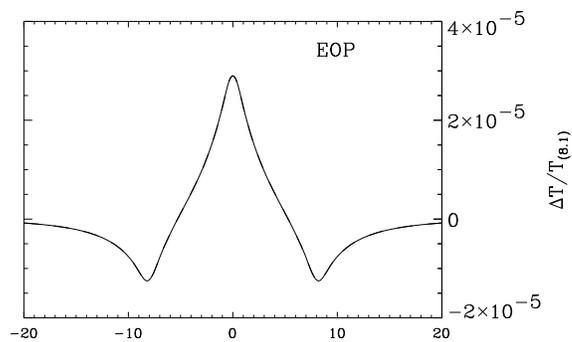
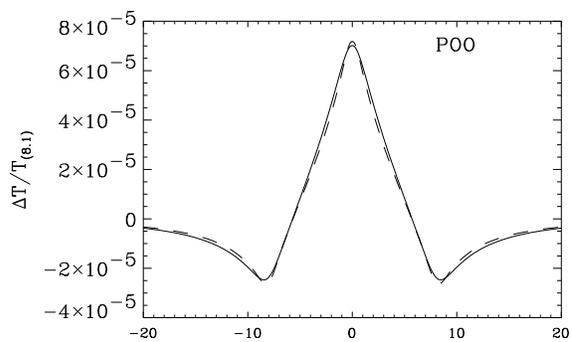
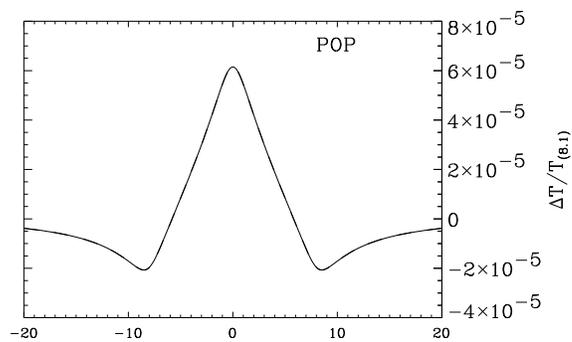
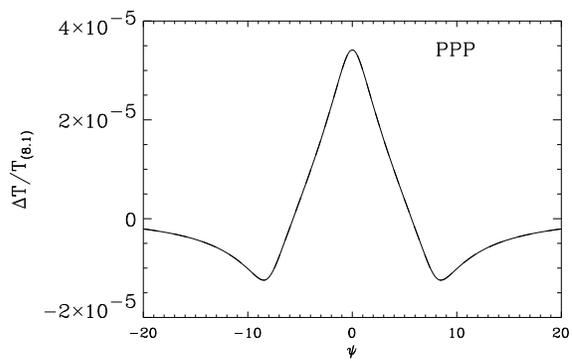
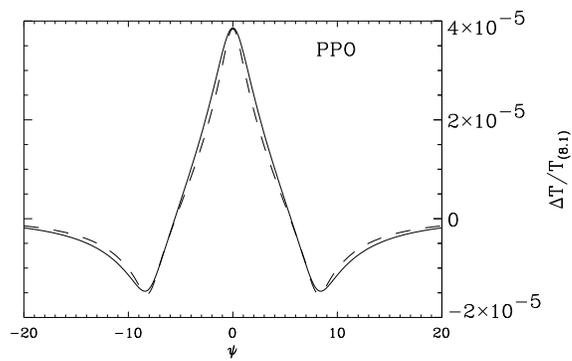
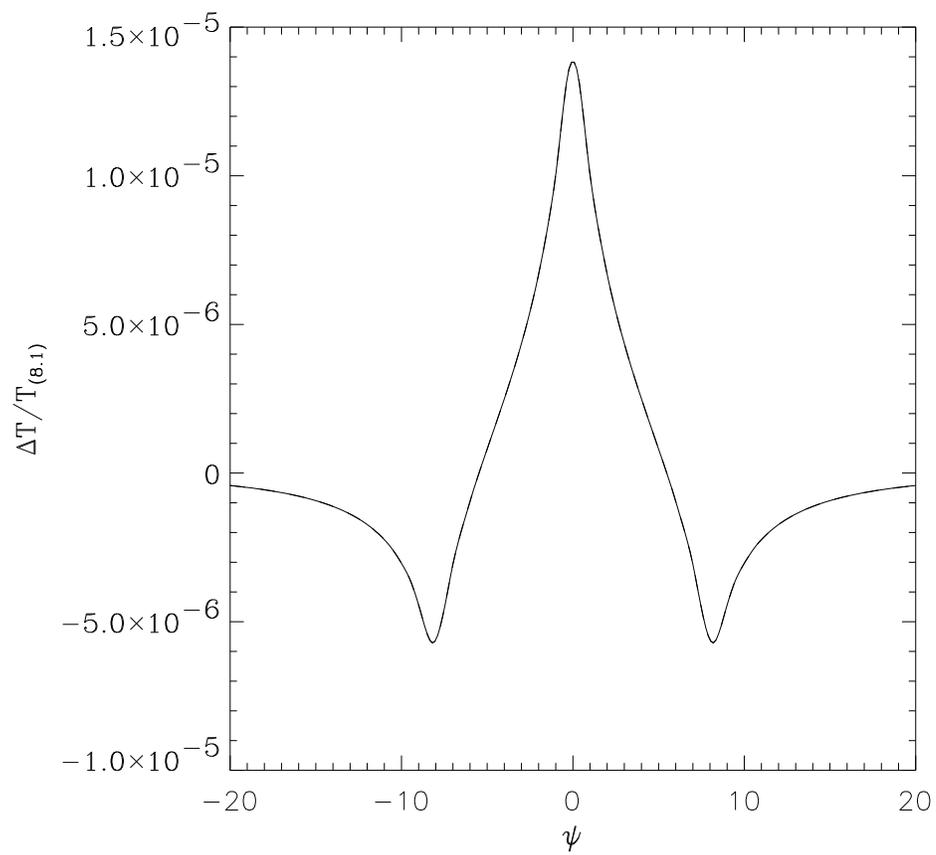
\end{document}